\documentclass[12pt]{article}

\usepackage{amssymb,amsmath,color,graphicx,dsfont,mathrsfs}

\newcommand{\spazioo}{{\cal C}^0(\Omega,\R)}

\newcommand{\HH}{{\mathcal H}}
\newcommand{\Haff}{H}
\newcommand{\hh}{{H}}
\newtheorem{theorem}{Theorem}
\newtheorem{corollary}[theorem]{Corollary}
\newtheorem{lemma}[theorem]{Lemma}
\newtheorem{proposition}[theorem]{Proposition}
\newtheorem{definition}[theorem]{Definition}
\newtheorem{remark}[theorem]{Remark}

\newenvironment{proof}{{\it Proof.}~~}{\hfill$\square$}

\renewcommand{\r}[1]{(\ref{#1})}

\newcommand{\Chi}{\chi}

\newcommand{\bt}{\begin{theorem}}
\newcommand{\et}{\end{theorem}}
\newcommand{\bl}{\begin{lemma}}
\newcommand{\el}{\end{lemma}}
\newcommand{\bp}{\begin{proposition}}
\newcommand{\ep}{\end{proposition}}
\newcommand{\bc}{\begin{corollary}}
\newcommand{\ec}{\end{corollary}}
\newcommand{\bdeff}{\begin{definition}}
\newcommand{\edeff}{\end{definition}}
\newcommand{\brem}{\begin{remark}}
\newcommand{\erem}{\end{remark}}
\newcommand{\bproof}{\begin{proof}}
\newcommand{\eproof}{\end{proof}}

\newcommand{\bi}{\begin{itemize}}
\newcommand{\iii}{\item}
\renewcommand{\i}{\item}
\newcommand{\ei}{\end{itemize}}
\newcommand{\bd}{\begin{description}}
\newcommand{\ed}{\end{description}}
\newcommand{\be}{\begin{enumerate}}
\newcommand{\ee}{\end{enumerate}}

\newcommand{\bqn}{\begin{eqnarray}}
\newcommand{\eqn}{\end{eqnarray}}
\newcommand{\eqnn}{\nonumber\end{eqnarray}}
\newcommand{\eqnl}[1]{\label{#1}\end{eqnarray}}

\newcommand{\ba}[1]{\begin{array}{#1}}
\newcommand{\ea}{\end{array}}

\newcommand{\R}{\mathbb{R}}

\newcommand{\N}{\mathbb{N}}
\newcommand{\C}{\mathbb{C}}
\newcommand{\Q}{\mathbb{Q}}
\newcommand{\lam}{\lambda}
\newcommand{\al}{\alpha}
\newcommand{\eps}{\varepsilon}

\newcommand{\Sn}{S^{2n-1}}

\newcommand{\Rep}{\mathfrak{X}}
\newcommand{\UU}{{U }}

\newcommand{\uu}{{u}}
\newcommand{\spazio}{{\rm sym}(n)}
\newcommand{\pa}{conically connected}

\title{
Approximate controllability, exact controllability, and conical eigenvalue intersections for quantum mechanical systems
}
\author{Ugo Boscain\thanks{Ugo Boscain,  CNRS, CMAP, \'Ecole Polytechnique, Palaiseau,   France,
\& Team GECO, INRIA Saclay,
{\tt ugo.boscain@polytechnique.edu}},
Jean-Paul Gauthier\thanks{Laboratoire LSIS, Universit\'e de Toulon, France  and Team GECO, INRIA Saclay, {\tt gauthier@univ-tln.fr}}, 
Francesco Rossi\thanks{Aix-Marseille Univ, LSIS, 13013, Marseille, France, {\tt francesco.rossi@lsis.org}},
Mario Sigalotti\thanks{INRIA Saclay, Team GECO \&
CMAP, \'Ecole Polytechnique, Palaiseau,   France,
 {\tt mario.sigalotti@inria.fr}}}

\begin{document}
\maketitle

\begin{abstract}
We study the controllability of 
a  closed control-affine quantum system driven by two or more external fields.
We provide a sufficient condition for controllability in terms of existence of conical intersections between eigenvalues of the Hamiltonian in dependence of the controls seen as parameters. 
Such spectral condition is structurally stable in the case of three controls or in the case of two controls when the  Hamiltonian is real. 
The spectral condition appears naturally in the adiabatic control framework and yields approximate  controllability in the infinite-dimensional case. 
In the finite-dimensional case 
it implies 
that the system is Lie-bracket generating when lifted to the group of unitary transformations, and in particular that it is exactly controllable.
Hence, Lie algebraic conditions are deduced from purely spectral properties.   

We conclude the article by proving that approximate and exact controllability are equivalent properties for general finite-dimensional quantum systems. 
\end{abstract}

\section{Introduction}

In this paper we  consider a closed quantum system of the form
\bqn
i\dot\psi(t)=H(u(t))\psi(t)=(H_0+u_1(t) H_1+\dots+u_m(t)H_m)\psi(t), 
\label{eq-0}
\eqn
where $\psi(\cdot)$ describes the state of the system evolving in the unit sphere $\mathcal{S}$ of a 
finite- or infinite-dimensional complex Hilbert space $\mathcal{H}$. 
The control $u(\cdot)=(u_1(\cdot),\dots,u_m(\cdot))$ takes values in a subset $U$ of $\R^m$  
and represents external fields. 
The Hamiltonian $H(u)$ is a self-adjoint operator on $\mathcal{H}$  for every $u\in U$.

System~\eqref{eq-0} is exactly (respectively, approximately) controllable if  every point 
of $\mathcal{S}$ can be steered   
to (respectively, steered arbitrarily close to) any other point of $\mathcal{S}$, 
by an admissible trajectory of \eqref{eq-0}.

When the dimension of $\mathcal{H}$ is finite, the exact controllability of \eqref{eq-0} has been characterized 
in \cite{albertini-dalessandro} in terms of the Lie algebra generated by $\{H(u)\mid u\in U\}$.  
In the infinite-dimensional case, if the controlled Hamiltonians $H_1$, \dots, $H_m$ are bounded, exact controllability can be ruled out 
by functional analysis arguments (\cite{bms,teismann,turinici}). 
Sufficient conditions for approximate controllability have been obtained by proving exact controllability of restrictions of \eqref{eq-0} to spaces where the controlled Hamiltonians are unbounded (\cite{beauchard,beauchard-coron,beauchard-laurent}). Other sufficient conditions for approximate controllability have been obtained by control-Lyapunov arguments (\cite{beauchard-nersesyan,mirrahimi,nersesyan,nersesyan-nersisyan}) and Lie--Galerkin techniques (\cite{weak,metalemma,weakly-coupled,periodic,mason}). 

Both in the finite- and the infinite-dimensional case, 
checking the above-mentioned controllability criteria  is not an easy task. 
Typical conditions require  that the eigenvalues of $H_0$ are non-resonant (e.g., all gaps are different or rationally independent) and that the controlled Hamiltonians ``sufficiently couple'' the eigenstates of $H_0$. 
Hence  many efforts were made to find easily checkable sufficient conditions for   controllability of (\ref{eq-0}). 

\begin{figure}
\begin{center}
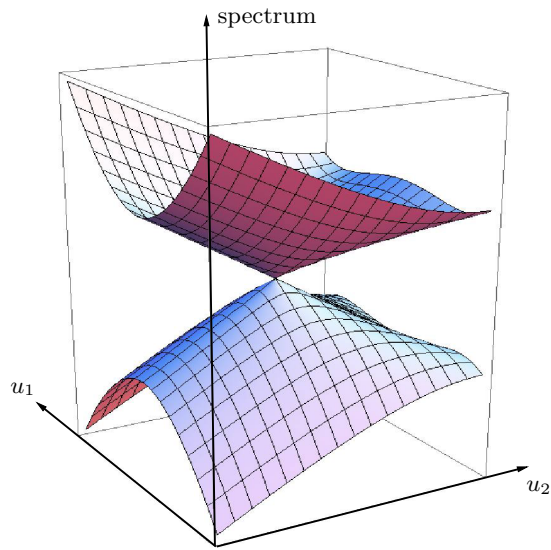
\caption{A conical intersection when $m=2$: the surfaces represent two eigenvalues of $H(u_1,u_2)$ as functions of $u_1$ and $u_2$.}\label{f-sing-singola}
\end{center}
\end{figure}

Notice that most of the conditions mentioned above are obtained for single-input systems ($m=1$). An alternative technique fully exploiting the multi-input framework 
uses adiabatic theory to obtain approximate descriptions of the evolution of \eqref{eq-0} for slowly varying control functions $u(\cdot)$ \cite{boscain-adami,adiabatic-TAC,
jauslin-guerin}. 
Adiabatic methods work when the spectrum exhibits eigenvalue intersections. In \cite{adiabatic-TAC}, in the case $m=2$, it is shown how to exploit the existence of \emph{conical intersections} (see Figure~\ref{f-sing-singola} and Definition~\ref{d-conical}) between every pair of subsequent eigenvalues to induce an approximate population transfer from any eigenstate to any other eigenstate or any nontrivial superposition of  eigenstates (without controlling the relative phases). This kind of partial 
controllability is named \emph{spread controllability} in \cite{adiabatic-TAC}.

In this paper we study the whole controllability implications of the conditions ensuring spread controllability, namely 
the existence of conical intersections between every pair of subsequent eigenvalues. A relevant advantage of these conditions is that they consist in qualitative structural properties of  the spectrum of $H(u)$ as a function of $u\in U$.  This 
might be useful when the explicit expression of the Hamiltonian is not known, but one has  information about its spectrum  (as it happens in many experimental situations).

In the following we say that \emph{the spectrum of $H(\cdot)$ is conically connected} if 
all eigenvalue intersections are conical, 
each pair  
of subsequent eigenvalues is connected by a conical intersections such that all other eigenvalues are simple (see Figure~\ref{f-sing-leggera}).
A notable property of conical connectedness is that it is a structurally stable property for $m=2$ (when restricted to real Hamiltonians) and for $m=3$. This structural stability dates back to the 1920s  (\cite{BF,vNW}) and is discussed in more details in Section~\ref{s-basic} (see Remark~\ref{gener}).

The main results of the paper about the relations between conically connected spectra and controllability  are the following: 
\begin{itemize}
\item if $\mathcal{H}$ is finite-dimensional and the spectrum of $H(\cdot)$ is conically connected then $\mathrm{Lie}\{H(u)\mid u\in U\}$ is equal to $\mbox{u}(n)$ 
if the trace of $H(u)$ is nonzero for some $u\in U$ or $\mbox{su}(n)$ otherwise. In particular \eqref{eq-0} is exactly controllable and the same is true for its lift in $\mbox{U}(n)$ or $\mbox{SU}(n)$; 
\item if $\mathcal{H}$ is infinite-dimensional and the spectrum of $H(\cdot)$ is conically connected then 
\eqref{eq-0} is 
approximately controllable. (For a counterpart of the finite-dimensional  lifted-system controllability, see Remark~\ref{r-simultaneous}.) 
\end{itemize}

Motivated by the exact/approximate dichotomy in the controllability of finite-/infinite-dimensional systems, we investigate in the last part of the paper the equivalence between exact and approximate controllability. 
We have already seen that exact controllability cannot hold when $\mathrm{dim}(\mathcal{H})=\infty$, since we 
assume $H_1$ and $H_2$ to be bounded. When $\mathrm{dim}(\mathcal{H})<\infty$ we prove that 
exact and approximate controllability are indeed equivalent, both for \eqref{eq-0} and its lift on $\mbox{U}(n)$ or $\mbox{SU}(n)$.
This last result holds in the more general setting where $H(u)$ depends on $u$ in a possibly nonlinear way.

\medskip
The structure of the paper is the following. In Section \ref{s-finito} we introduce the basic definitions related to controllability and conical intersections and we prove the finite-dimensional exact controllability  of a system exhibiting a conically connected spectrum and of its lift in $\mbox{U}(n)$  or  $\mbox{SU}(n)$ (Theorem~\ref{t-main}). In Section \ref{s-infinite} we prove that an infinite-dimensional system having a conically connected spectrum is approximately controllable (Theorem~\ref{t-duduk}). Finally, in Section \ref{s-equivalence} we prove the equivalence between approximate and exact controllability for finite-dimensional closed  quantum mechanical systems.

\section{Conical intersections and exact controllability in finite dimension}
\label{s-finito}\label{s-finite}

\subsection{Basic definitions and facts}
\label{s-basic}
 In this section 
 we introduce some definitions and recall some basic facts about 
 control systems evolving on  finite-dimensional 
 manifolds. 

We first define approximate and exact controllability for a 
smooth control system 
$$
 \dot q(t)=f(q(t),u(t)) \eqno{(\Sigma)}
$$ 
defined on a connected manifold $M$ with controls $u(\cdot)$ taking values in $U\subset \R^m$.

\bdeff 

\mbox{}

\bi

\i The {\em reachable set} $\mathcal{A}_{q_0}$ from a point $q_0\in M$ for  $(\Sigma)$ is the set of points $q_1\in M$ such that there exist a time $T\geq 0$ and  a $L^\infty$ control $u:[0,T]\to U$ for which the solution of the Cauchy problem  $\dot q(t)=f(q(t), u(t))$ starting from $q(0)=q_0$ is well defined on $[0,T]$  and satisfies $q(T)=q_1$. 

\i The system $(\Sigma)$  is said to be {\em exactly controllable} if for every $q_0\in M$  we have $\mathcal{A}_{q_0}=M$.

\i The system $(\Sigma)$  is said to be {\em approximately controllable} if for every $q_0\in M$  we have that $\mathcal{A}_{q_0}$ is dense in $M$. 
\ei
\edeff

A relevant class of control systems for our discussion is given by right-invariant control systems on Lie groups, namely, systems for which $M$ is a connected Lie group and each vector field $f(\cdot,u)$, $u\in U$, is right-invariant. 

Lemma \ref{l-compact} below is a classical result concerning right-invariant control systems on compact Lie groups (see, e.g., \cite{elass} and \cite[p.~155]{djurd}).

\bdeff\label{d-orbit}
Let $(\Sigma)$ be a right-invariant control system and  denote by $e$ the identity of the group $M$.
Let $\mathrm{Lie}\{f(e,u)\mid u\in U\}$ be the {\rm Lie algebra generated by} $\{f(e,u)\mid u\in U\}$, i.e., the smallest subalgebra of the Lie algebra 
of $M$ containing $\{f(e,u)\mid u\in U\}$. 
 The {\rm orbit} 
$G$
of $(\Sigma)$ is the connected subgroup of $M$ whose Lie algebra is $\mathrm{Lie}\{f(e,u)\mid u\in U\}$.
\edeff

\bl\label{l-compact} Let $M$ be a connected compact Lie group and consider a right-invariant control system $(\Sigma)$ on $M$.
The following conditions are equivalent: 
\bi
\i $(\Sigma)$  is exactly controllable;
\i the orbit $G$ of $(\Sigma)$ is equal to $M$; 
\i $\mathrm{Lie}\{f(e,u)\mid u\in U\}$ is the Lie algebra of $M$.
\ei
\el
The last condition is usually referred to as the {\em Lie-bracket generating condition}.

A general controlled closed quantum system evolving in a 
 finite-dimensional Hilbert space can be written as
\bqn\label{gqs}
i\dot\psi(t)=\hh(u(t))\psi(t),
\eqn
where 
$\psi:[0,T]\to S^{2n-1}\subset \C^n$ denotes the state of the system and $\hh(u)$ is a Hermitian matrix smoothly depending on $u\in U\subset \R^m$. From now on let us take $n\geq 2$, otherwise the controllability problem is trivial.

Naturally associated with \eqref{gqs} is its lift on the unitary group $\mbox{U}(n)$,
\bqn\label{gqs-lift}
i\dot g(t)=\hh(u(t))g(t),
\eqn
which is right-invariant and permits to write the solution $\psi(\cdot)$ of \eqref{gqs} starting from $\psi_0$ as  
 $\psi(t)=g(t)\psi_0$ where $g(\cdot)$ is the solution of \eqref{gqs-lift} starting from the identity.

Lemma~\ref{l-compact} implies that \eqref{gqs-lift}
is controllable in $\mbox{U}(n)$ if and only if the Lie algebra generated by $\{i \hh(u)\mid u\in U\}$ is equal to $\mbox{u}(n)$. 
 If the trace of each matrix $\hh(u)$, $u\in U$, is zero, then \eqref{gqs-lift} is well posed in $\mbox{SU}(n)$ and its exact controllability in $\mbox{SU}(n)$ is equivalent to the condition 
$\mathrm{Lie}\{i \hh(u)\mid u\in U\}= \mbox{su}(n)$.

In order to deduce the controllability properties of \eqref{gqs} from those of \eqref{gqs-lift} one 
has to turn towards the classification of transitive actions of subgroups of $\mbox{U}(n)$ onto $S^{2n-1}\subset \C^n$. 
As a consequence, system (\ref{gqs}) 
is exactly controllable if and only if 
 \begin{equation}\label{daL}
\mathrm{Lie}\{i \hh(u)\mid u\in U\}\supseteq\left\{\ba{l}\mbox{su}(n)\mbox{ if $n$ is odd }\\ 
\mbox{an algebra conjugate to }\mbox{sp}(n/2) \mbox{ if $n$ is even.}
\ea\right.
\end{equation}
(See \cite{dalessandro}.)

Of special interests for this paper are  closed control-affine quantum system driven by $m$ external fields, satisfying 
the following assumption:
 \bi
\iii[ {\bf (A)}] 
Let $m\geq 2$ and $U$ be an open and connected  subset 
of  $\R^m$. 
  We assume that $H(\cdot)$ is control-affine, i.e., it has the form 
  $$H(u)=H_0+u_1 H_1+\cdots+u_mH_m.$$
\ei

In the following, under assumption {\bf (A)}, we focus on the controllability of the system
\bqn
i\dot\psi(t)=(H_0+u_1(t) H_1+\cdots+u_m(t) H_m)\psi(t),\qquad \psi(t)\in S^{2n-1},
\label{eq-1}
\eqn
and its lift
\bqn
i\dot g(t)=(H_0+u_1(t) H_1+\cdots+u_m(t) H_m)g(t),\qquad g(t)\in \mbox{U}(n).
\label{eq-group}
\eqn

\begin{remark}
Let us briefly discuss the role of the assumptions listed in hypotheses {\bf (A)}. 
The affine structure of $H$ with respect to the control is natural in quantum control (\cite{dalessandro}) and 
allows the application of the controllability criteria  we are using in the following (see Proposition~\ref{marcoefinito}). 
Moreover, the connectedness of $U$ is required in order to apply adiabatic techniques in the whole set of control parameters. 
\end{remark}


%
%


A crucial hypothesis that we shall use to prove exact controllability of \eqref{eq-group} (and hence, in particular, of \eqref{eq-1}) 
 is the existence of conical intersections (in the space of controls) between consecutive energy levels, and the fact that  these conical intersections occur at distinct points in the space of controls. More precisely:
\bdeff\label{d-conical}
Let  {\bf (A)} be satisfied.
Let $\Sigma(\uu)=\{\lambda_1(\uu) ,\ldots,\lambda_n(\uu)\}$ be the spectrum of $\Haff(\uu)$, where the eigenvalues $\lambda_1(\uu)\leq \cdots \leq\lambda_n(\uu)$ are counted according to their multiplicities.
%
 We say that $\bar\uu\in\UU$ is a \emph{conical intersection} between the eigenvalues $\lam_j$ and $\lam_{j+1}$ if $\lam_{j}(\bar \uu) = \lam_{j+1}(\bar \uu)$ has multiplicity two and there exists a constant $c>0$ such that for any unit vector ${v}\in \R^m$ and $t>0$ small enough we have 
\begin{equation} \label{formcono}
\lam_{j+1}(\bar \uu+t{v})-\lam_{j}(\bar \uu+t{v}) > ct\,.\end{equation}
\edeff

See Figure~\ref{f-sing-singola} for the picture of a conical intersection.
Notice that the hypothesis $m\geq 2$ guarantees that conical intersections do not disconnect $U$. This 
is crucial in the arguments below (see, in particular, Lemma~\ref{l-gaulthier}.)

\begin{remark}\label{gener}
Conical intersections are not pathological
phenomena. On the contrary, they happen to be generic for  $m=3$ or for $m=2$, when restricted to real Hamiltonians, 
in the following sense.

Let us first consider the case $m=2$. Let $\spazio$ be the set of all $n\times n$ symmetric real matrices.
Then, generically with respect to the pair $(H_1,H_2)$ in
$\spazio\times \spazio$  (i.e., for all $(H_1,H_2)$ in an open and dense subset of $\spazio\times \spazio$),
for each $\uu=(u_1,u_2) \in \mathbb{R}^2$ and $\lambda\in \R$ such that $\lambda$ is a multiple
eigenvalue of
$H_0+u_1 H_1+u_2 H_2$, the eigenvalue intersection $\uu$ is conical. 
Moreover, each
conical intersection $\uu$ is structurally stable, in the sense
that  small perturbations of $H_0$, $H_1$ and $H_2$ give rise, in a neighborhood of $\uu$, to conical
intersections for the perturbed $\Haff$. See Section~\ref{s-infinite} for a version of this result in infinite dimension and \cite{adiabatic-TAC} for more details.

In the case $m=3$, let $\mathrm{Herm}(n)$ be the space of $n\times n$ Hermitian matrices. Then, generically with respect to the triple $(H_1,H_2,H_3)$ in $\mathrm{Herm}(n)^3$, 
for each $\uu=(u_1,u_2,u_3) \in \mathbb{R}^3$ and $\lambda\in \R$ such that $\lambda$ is a multiple
eigenvalue of
$H_0+u_1 H_1+u_2 H_2+u_3 H_3$, the eigenvalue intersection $\uu$ is conical. Structural stability also holds, in the same sense as above.
See \cite{preprint-z} for more details and a discussion on the infinite-dimensional counterpart of these properties.  
\end{remark}

The following definition identifies the Hamiltonians for which we can guarantee exact controllability  from qualitative properties of their spectra. Roughly speaking  we require all their eigenvalues to be connected by conical intersections and the conical intersections to occur at different points in the space of controls.
\bdeff\label{d-conicallyconnected}
Let  {\bf (A)} be satisfied. We say that the spectrum $\Sigma(\cdot)$ of  $\Haff(\cdot)$ is {\em \pa} if {all eigenvalue intersections are conical} and for every  $j=1,\ldots,n-1$, 
 there exists a conical  intersection $\bar \uu_j\in\UU$ between the eigenvalues $\lam_j,\lam_{j+1}$, with $\lambda_l(\bar\uu_j)$ simple if $l\neq j,j+1$.
\edeff

\begin{figure}
\begin{center}
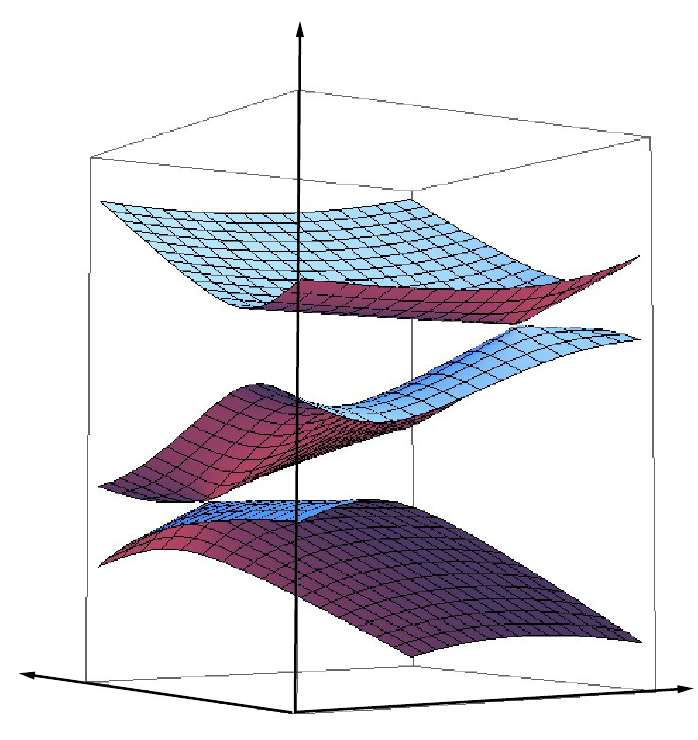
\caption{A  \pa\ spectrum in the case $m=2$.}\label{f-sing-leggera}
\end{center}
\end{figure}

See Figure  \ref{f-sing-leggera} for a \pa\ spectrum. 

\subsection{Conical connectedness  implies exact controllability}

The main result of Section \ref{s-finito} is the following theorem.
\bt
\label{t-main}
Let  {\bf (A)} be satisfied and assume that the   spectrum $\Sigma(\cdot)$ of $H(\cdot)$ is \pa.  Then  
the Lie algebra generated by $\{i H(u)\mid u\in U\}$ is either $\mbox{u}(n)$ or $\mbox{su}(n)$ (in the case $H_0,\dots,H_m\in\mbox{su}(n)$). 
Hence,
system (\ref{eq-group}) is either exactly controllable in $\mbox{U}(n)$ or well-posed and exactly controllable in $\mbox{SU}(n)$.
\et


The proof of the theorem is based on the following lemma.

\bl
\label{l-gaulthier}
Let  {\bf (A)} be satisfied and assume that the  spectrum $\Sigma(\cdot)$ of $H(\cdot)$ is \pa.  
Then there exists $\bar\UU\subset \UU$ which is dense and with zero-measure complement in $\UU$ such that 
if $\sum_{j=1}^n\al_j \lam_j(\bar \uu)=0$ with 
 $(\al_1,\ldots,\al_n)\in\Q^n$ and  $\bar \uu\in\bar\UU$ then $\al_1=\al_2=\dots=\al_n$.
\el

\newcommand{\Us}{\UU_{\mathrm{simple}}}
\proof 
For every $\al=(\al_1,\ldots,\al_n)\in\Q^n$ 
define
$$
\UU_\alpha=\{\uu\in\UU\mid \sum_{j=1}^n \alpha_j\lambda_j(\uu)=0 \}.
$$
Let $\bar U$ be the complement in $U$ of the union of  all $U_\alpha$ such that $\alpha_j\neq\alpha_k$  for some $j,k\in\{1,\ldots,m \}$.

Notice 
that, by definition of conical intersection and since $m\geq 2$,  
$\{\uu\in \UU\mid \Sigma(\uu)\mbox{ is simple}\}$ is connected. 
Thanks to the analyticity of the spectrum in $\{\uu\in \UU\mid \Sigma(\uu)\mbox{ is simple}\}$, either $\UU_\alpha=\UU$
or $\UU_\alpha$
has empty interior.  The proof is completed by showing that if $\UU_\alpha=\UU$ then $\alpha_1=\dots=\alpha_n$. Indeed, in this case $U_\alpha$ has empty interior for every $\alpha$ such that  $\alpha_j\neq\alpha_k$ for some $j,k\in\{1,\ldots,m \}$ and the countable union of subsets of $\R^m$ with empty interior and zero measure has empty interior and zero measure. 

Assume that $\UU_\alpha=\UU$. 
Consider $j\in\{1,\ldots,n-1\}$ and an
 analytic path $\gamma:\R\to\UU$ such that $\gamma(0)=\bar \uu_j$,  $\dot \gamma(0)\neq 0$, where 
  $\bar \uu_j\in\UU$ is a conical intersection between  the eigenvalues $\lam_j,$ and $\lam_{j+1}$, with $\lambda_l(\bar\uu_j)$ simple if $l\neq j,j+1$.

Since $\UU_\alpha=\UU$, we have for every $t\in\R$,
$$\sum_{l=1}^n \alpha_l\lam_l(\gamma(t))=0.$$ 
By analytic dependence of the spectrum along $\gamma$ in a neighbourhood of $\gamma(0)$ \cite{rellich}, the functions
$$
t\mapsto\left\{\ba{ll} \lam_j (\gamma(t))&\mbox{ if } t<0 \\  \lam_{j+1} (\gamma(t))&\mbox{ if } t\geq0,\ea\right.\qquad t\mapsto\left\{\ba{ll} \lam_{j+1} (\gamma(t))&\mbox{ if } t<0 \\  \lam_{j} (\gamma(t))&\mbox{ if } t\geq0,\ea\right.
$$
and $t\mapsto \lambda_l(\gamma(t))$, $l\ne j,j+1$,  are analytic in a neighborhood of $0$.  
Hence, 
$$\alpha_{j+1}\lam_j(\gamma(t))+ \alpha_{j}\lam_{j+1}(\gamma(t))+\sum_{l\neq j,j+1} \alpha_l\lam_l(\gamma(t))=0$$ 
for $t$ in  a neighborhood of $0$. 
Then
$$
 (\alpha_j-\alpha_{j+1}) (\lam_j(\gamma(t))-\lam_{j+1}(\gamma(t)))=0
$$
for $t$ in  a neighborhood of $0$. By definition of conical intersection it must be $\alpha_j=\alpha_{j+1}$. Since $j$ is arbitrary, 
we deduce that $\alpha_1=\dots=\alpha_n$ 
concluding the proof. \hfill$\Box$
\begin{remark}
The lemma fails to hold if $m=1$,  i.e., for single input systems.
Consider for instance $n=3$, $H_0=\mathrm{diag}(0,1,2)$ and $H_1=\mathrm{diag}(1,1,0)$.
Then the eigenvalues of $H(u)$ are $u,u+1$ and $2$. The spectrum is conically connected, but clearly $\bar U=\emptyset$. 

Notice that $\mathrm{Lie}(i H_0,i H_1)$ is made only by diagonal matrices and therefore $\{i H_0,i H_1\}$ does not generate $\mbox{u}(n)$. Hence, this example also shows that Theorem~\ref{t-main} 
does not hold if we remove the hypothesis $m\geq 2$. 
\end{remark}

The proof of Theorem~\ref{t-main} is based on the following adaptation of a controllability criteria for single-input quantum control systems appeared in \cite[Proposition 3.1]{weak}. The proof can be obtained following exactly the same arguments as in \cite{weak}.

\bp\label{marcoefinito}
Let 
$A_0,A_1,\dots,A_m$ be skew-Hermitian $n\times n$ matrices. 
Denote by 
$\lam_1,\dots,\lam_n$ the eigenvalues of
 $A_0$, repeated according to their multiplicities and let $\phi_1,\dots,\phi_n$ be an orthonormal basis of associated eigenvectors.
 Let 
 $$S_0=\{(j,k)\in  \{1,\dots,n\}^2 \mid \mbox{$\exists\; l\in\{1,\dots,m\}$ such that } \langle \phi_j,A_l \phi_k\rangle\ne 0\}.$$
 
%
%

Assume that there exists $S\subseteq S_0$ such that
the graph having $1,\dots,n$ as nodes and $S$ as set of edges is connected. Assume,  moreover, that 
 for every $(j,k)\in S$ and $(r,s)\in S_0\setminus\{(j,k)\}$ we have   $\lam_j-\lam_k\ne \lam_r-\lam_s$.
Then $\mbox{Lie}(A_0,\dots,A_m)=\mbox{su}(n)$ if $A_0,\dots,A_m\in \mbox{su}(n)$ and 
$\mbox{Lie}(A_0,\dots,A_m)=\mbox{u}(n)$ otherwise.
\ep

\noindent {\it Proof of Theorem~\ref{t-main}.}
Applying Lemma~\ref{l-gaulthier} we deduce the existence of $\uu_0\in \UU$ such that 
if $\sum_{j=1}^n\al_j \lam_j(\uu_0)= 0$ with
 $(\al_1,\ldots,\al_n)\in\Q^n$  then $\al_1=\dots=\al_n$.
In particular, the spectrum of $\Haff(\uu_0)$ is simple and two spectral gaps $\lambda_j(\uu_0)-\lambda_k(\uu_0)$ and $\lambda_r(\uu_0)-\lambda_s(\uu_0)$ are different if $(j,k)\ne (r,s)$ and $j\ne k$, $r\ne s$.
Let $\phi_1,\dots,\phi_n$ be an orthonormal basis of  eigenvectors of $\Haff(\uu_0)$. 

 Let us conclude the proof by applying Proposition~\ref{marcoefinito} to $A_0=i \Haff(\uu_0)$, $A_j=i H_j$ for $j=1,\dots,m$: to this purpose, we are left to prove that 
 the graph having $1,\dots,n$ as nodes and 
$$S_0 =\{(j,k)\in  \{1,\dots,n\}^2 \mid \langle \phi_j,H_l \phi_k\rangle\ne 0\mbox{ for some }l=1,\dots,m\}$$
as set of edges is connected.

Assume by contradiction that such graph is not connected. Then there exists a proper subspace $V$ of $\C^n$ generated by eigenvectors of $\Haff(\uu_0)$ which is invariant for the evolution of \eqref{eq-1}. Without loss of generality $V=\mathrm{span}\{\phi_1,\dots,\phi_r\}$ with $r<n$. 

Since the spectrum is conically connected, we can apply \cite[Corollary 2.5]{teufel}  and deduce that there exists an admissible trajectory of \eqref{eq-1} steering $\phi_1$ to an arbitrary small neighbourhood of $\{e^{i\theta} \phi_n\mid \theta\in \R\}$.   (See also \cite[Proposition 3.4]{adiabatic-TAC} for a rephrasing in control terms of  \cite[Corollary 2.5]{teufel}, which deals with general adiabatic evolutions through conical intersections.  The result is stated in \cite{adiabatic-TAC} in the case $m=2$ for symmetric Hamiltonians but actualy holds in the general case.)   
The contradiction is reached, since $V\cap \{e^{i\theta} \phi_n\mid \theta\in \R\}=\emptyset$.
\hfill$\Box$

\section{Conical intersections and approximate controllability in infinite dimension}\label{s-infinite}

In this section we  extend the controllability analysis of the previous section to systems of the form \eqref{eq-1} evolving in infinite-dimensional spaces. 

Consider a separable infinite-dimensional complex Hilbert space $\HH$. 
In this section
 we 
make the following assumption:

 \bi
\iii[ {\bf (A$^\infty$)}] 
 Let $m\geq 2$ and 
 $U$ be an open and connected  subset 
of  $\R^m$. 
%
Assume that the Hamiltonian $H(\cdot)$ has the form
\[
\Haff(\uu) = H_0 + u_1 H_1 +\cdots+u_m H_m,\quad \uu = (u_1,\dots, u_m) \in \UU,
\]
where  
 $H_0,\dots,H_m$ are self-adjoint operators on $\HH$, with $H_0$ bounded from below and $H_1,\dots,H_m$ bounded. 
\ei

With a Hamiltonian $H(\cdot)$ as in assumption {\bf (A$^\infty$)} we can associate the control system 
\bqn
i\dot\psi(t)=(H_0+u_1(t) H_1+\cdots +u_m(t) H_m)\psi(t),\qquad \psi(t)\in {\cal S},
\label{eq-2}
\eqn
where ${\cal S}$ is the unit sphere of $\HH$.

Existence of solutions of \eqref{eq-2} 
for $u$ of class $L^\infty$ and $H_1,\dots,H_m$ bounded is classical (see \cite{pazy-book}).

%

A typical case for which {\bf (A$^\infty$)} is satisfied is when $H_0=-\Delta  + V$, where $\Delta$ is the Laplacian on a domain $\Omega\subset \R^d$
 (with suitable boundary conditions if $\Omega\ne \R^d$),  
$V$ is a regular enough real-valued potential bounded from below, ${\cal H}=L^2(\Omega,\C)$, 
and $H_1, \dots,H_m$ are  multiplication operators by $L^\infty$ real-valued functions. 

%

\subsection{Conical connectedness implies approximate controllability in infinite dimension}

The main technical assumption of this section is the following.
 \bi
\iii[ {\bf (B)}] 
The spectrum of $H_0$ is discrete without accumulation points and each eigenvalue has finite multiplicity. 
\ei

Under assumptions {\bf (A$^\infty$)} and {\bf (B)}
the spectrum of $\Haff(\uu)$, $\uu\in\UU$, with eigenvalues repeated according to their multiplicities, can be described by $\Sigma^\infty(\uu)=\{\lambda_j(\uu)\}_{j\in \N}$ with $\lambda_j(\uu)\leq \lambda_{j+1}(\uu)$ for every $j\in \N$ and each $\lambda_j(\cdot)$ continuos on $\UU$. 
In analogy with Definition~\ref{d-conicallyconnected}, we say that $\Sigma(\cdot)$ is \emph{conically connected} if all eigenvalue intersections $\lambda_j=\lambda_{j+1}$, $j\in\N$, are conical (the definition of conical intersection extends trivially to this case) and for every  $j\in\N$
 there exists a conical  intersection $\bar \uu_j\in\UU$ between the eigenvalues $\lam_j,\lam_{j+1}$, with $\lambda_l(\bar\uu_j)$ simple if $l\neq j,j+1$.

\begin{remark}\label{genr-infty}
Recall from \cite{adiabatic-TAC} that conical intersections are generic in the case $m=2$ 
in the reference case where $\mathcal{H}=L^2(\Omega,\C)$,
$H_0=-\Delta+V_0:D(H_0)=H^2(\Omega,\C)\cap H^1_0(\Omega,\C)\to
L^2(\Omega,\C)$, $H_1=V_1$, $H_2=V_2$, with $\Omega$ a bounded domain
of $\R^d$ 
and $V_j\in \spazioo$ for $j=0,1,2$.
Indeed, generically with respect to the pair $(V_1,V_2)$ in
$\spazioo\times \spazioo$ (i.e., for all $(V_1,V_2)$ in a
countable intersection of open and dense  subsets of $\spazioo\times \spazioo$),
for each $\uu \in \mathbb{R}^2$ and $\lambda\in \R$ such that $\lambda$ is a multiple
eigenvalue of
$H_0+u_1 H_1+u_2 H_2$, the eigenvalue intersection $\uu$ is conical.
Moreover, each
conical intersection $\uu$ is structurally stable, in the sense
that  small perturbations of $V_0$, $V_1$ and $V_2$ give rise, in a neighbourhood of $\uu$, to conical
intersections for the perturbed $H$.
\end{remark}

The main purpose of this section is to extend Theorem~\ref{t-main} 
to the infinite-dimensional case, as follows.

\bt
\label{t-duduk} 
Let  hypotheses {\bf (A$^\infty$)} and {\bf (B)} be satisfied.
If the spectrum $\Sigma(\cdot)$  is \pa\ then (\ref{eq-2}) is  approximately controllable. 
\et

The proof of Theorem~\ref{t-duduk} follows the same 
pattern as the one of Theorem~\ref{t-main}. 
The first step is the following straightforward generalisation of Lemma~\ref{l-gaulthier}.

\bl
\label{l-gaulthier-infty}
Let  hypotheses {\bf (A$^\infty$)} and {\bf (B)} be satisfied and assume that the  spectrum $\Sigma(\cdot)$ is \pa.  
Then there exists $\bar\UU\subset \UU$ which is dense and with zero-measure complement in $\UU$ such that for each $N\in \N$,  $\sum_{j=1}^N\al_j \lam_j(\bar \uu)=0$ with 
 $(\al_1,\ldots,\al_N)\in\Q^N$ and  $\bar\uu\in\bar\UU$ implies $\al_1=\al_2=\dots=\al_N=0$.
\el

In particular the spectrum of $\Haff(\bar\uu)$ for $\bar\uu\in \bar \UU$ as in Lemma~\ref{l-gaulthier-infty} is such that two spectral gaps
$\lambda_k(\bar \uu)-\lambda_j(\bar \uu)$ and $\lambda_r(\bar \uu)-\lambda_s(\bar \uu)$ are different if $(k,j)\ne (r,s)$ and $k\ne j$, $r\ne s$.


In the infinite-dimensional case, the role of Proposition~\ref{marcoefinito} is played by the following corollary of \cite[Theorem 2.6]{weak}.

\bp\label{corollary-metalemma}
Let  hypotheses {\bf (A$^\infty$)} and {\bf (B)} be satisfied.
Assume that there exists $\bar \uu\in \UU$ such that  
$\lambda_k(\bar \uu)-\lambda_j(\bar \uu)\ne \lambda_r(\bar \uu)-\lambda_s(\bar \uu)$ if $(k,j)\ne (r,s)$, $(k,j),(r,s)\in\N^2\setminus\{(l,l)\mid l\in\N\}$. Denote by $(\phi_j(\bar \uu))_{j\in\N}$ a Hilbert basis of eigenvectors of $\Haff(\bar \uu)$ and let
 $$S=\{(j,k)\in  \N^2 \mid  \langle \phi_j(\bar \uu),H_l \phi_k(\bar \uu)\rangle\ne 0 \mbox{ for some }l=1,\dots,m\}.$$
 
%
%
If the graph having $\N$ as set of nodes and $S$ as set of edges is connected then \eqref{eq-2} is approximately controllable in $\mathcal{S}$.
\ep

The proof of Theorem~\ref{t-duduk} is then concluded as
follows: Lemma~\ref{l-gaulthier-infty} 
 guarantees the existence of $\bar \uu$ such that the spectral gaps of $\Sigma(\bar \uu)$ are all  different; this allows to deduce the conclusion from Proposition~\ref{corollary-metalemma} provided that 
no proper linear subspace of $\mathcal{H}$ spanned by eigenvectors of $\Haff(\bar\uu)$ is  invariant for \eqref{eq-2}.
As in the finite-dimensional case, this can be be proved by adiabatic  methods, deducing from \cite[Corollary 2.5]{teufel} (or  \cite[Proposition 3.4]{adiabatic-TAC}) that for every pair of eigenvectors of $\Haff(\bar \uu)$ there exists and admissible trajectory of \eqref{eq-2} connecting them with arbitrary precision.

\begin{remark}\label{r-simultaneous}
Following \cite{metalemma}, a stronger version of 
Proposition~\ref{corollary-metalemma}, and hence of Theorem~\ref{t-duduk}, could be stated, namely: under the same hypotheses, for every $l\in \N$, $\psi_1,\dots,\psi_l\in\mathcal{S}$, $\eps>0$,   and every unitary  transformation $\Upsilon$ of $\mathcal{H}$, there exists a  control function $\uu:[0,T]\to \UU$ such that, for every $j=1,\dots,l$ the solution of \eqref{eq-2} having 
$\psi_j$ as initial conditions arrives in a $\eps$-neighborhood of $\Upsilon(\psi_j)$ at time $T$. 
Notice that this is the natural counterpart of controllability of the lift of \eqref{eq-1} in the group of unitary transformations proved in Section~\ref{s-finite}. 
\end{remark}

\section{Equivalence between exact and approximate controllability for finite-dimensional systems}
\label{s-equivalence}

In the previous sections we have 
seen several sufficient conditions for controllability, which is exact in the finite-dimensional case and approximate  in the  infinite-dimensional one. 


Our aim is to show that in the finite-dimensional case approximate controllability always yields exact controllability for systems of the type
\bqn\label{gqs-2}
i\dot\psi(t)=\hh(u(t))\psi(t),\qquad \psi(t)\in S^{2n-1},\ u(t)\in U\subset\R^m,  
\eqn
or
\bqn\label{gqs-lift-2}
i\dot g(t)=\hh(u(t))g(t),\qquad g(t)\in \mathscr{G},\ u(t)\in U\subset\R^m,
\eqn
where $\mathscr{G}$ denotes the group $\mbox{SU}(n)$ if the trace of $\hh(u)$ is zero for every $u\in U$ and $\mbox{U}(n)$ otherwise. 

More precisely, we have the following.
\bt
\label{t-2}
System (\ref{gqs-2}) is approximately controllable if and only it is exactly controllable. The same holds for system \eqref{gqs-lift-2}.
\et

\subsection{Remarks on Theorem~\ref{t-2}}

The proof of Theorem~\ref{t-2} is based on some results in representation theory,  recalled in the following section.

The statement of Theorem~\ref{t-2} for the lifted problem in $\mbox{SU}(n)$  is folklore. 
Indeed, the proof follows from the following 1942 result by Smith \cite[note on p.~312]{smith}, as detailed below.

\bt[\cite{smith}]\label{t-gr}
If a dense subgroup $\hat G$ of a simple Lie group $G$ of dimension larger than $1$ contains an analytic arc, then $\hat G=G$.
\et

\noindent
{\it Proof of Theorem~\ref{t-2} in the case $\mathscr{G}=\mbox{SU}(n)$.} 
Let \r{gqs-lift-2} be approximately controllable in $\mbox{SU}(n)$. Then, the orbit from the identity is a dense subgroup $\hat{\mathscr{G}}$ of $\mbox{SU}(n)$. Any trajectory of \r{gqs-lift-2} with constant $u$ is an analytic arc, contained in $\hat{ \mathscr{G}}$. Then $\hat{\mathscr{G}}=\mbox{SU}(n)$, i.e., the orbit is the whole group.  Lemma \ref{l-compact} yields that the accessible set coincides with $\mathscr{G}$, i.e., that  system \r{gqs-lift-2} is exactly controllable. \hfill $\Box$

Notice that the argument does not apply for $\mathscr{G}=\mbox{U}(n)$, since $\mbox{U}(n)$ is not simple. 
Moreover, the equivalence between approximate and exact controllability on the sphere does not follow from the 
result on the lifted system. It is well-known, indeed, 
that approximate/exact controllability on the group and on the sphere are not equivalent since, as already recalled, 
if the Lie algebra generated by $\{i H(u)\mid u\in U\}$ is equal to $\mbox{sp}(n/2)$ then \eqref{gqs-2} is exactly controllable, while \eqref{gqs-lift-2} is not (even approximately).

\subsection{Some facts  from group-representation theory}
\label{s-lie}

\newcommand{\hf}{h}

In this section, we recall the two basic main facts from representation theory that are needed in order to prove Theorem~\ref{t-2}.
We consider a finite-dimensional representation of a Lie group $G$,  $\Rep:~G\to L(\hf)$, where $\hf$ is a finite dimensional complex Hilbert space and  $L(\hf)$ denotes 
the space of endomorphisms of $\hf$. 

Theorem \ref{t-dix} below is stated by  Dixmier in \cite{dixm}. We need it for Lie groups, although it holds more generally for locally compact topological groups.  

We recall that the intersection of the kernels of all unitary irreducible finite-dimensional representations of a group $G$ is a subgroup of $G$. Then,  $G$  is said to be \emph{injectable in a compact group}\footnote{The definition given here is not the most  natural, since injectability in a  compact group is related to the notion of \emph{compact group associated with a topological group} that is defined via an universal property: For each topological group $G$ there exists a compact group $\Sigma$ and a continuous morphism $\alpha:G\to \Sigma$ such that for any compact group $\Sigma'$ and continuous morphism $\al':G\to \Sigma'$ it exists a continuous morphism $\beta:\Sigma\to\Sigma'$ such that $\alpha'=\beta\circ \alpha$. We give here only the definition that fits better with our purposes. For such beautiful theory, see  \cite[16.4]{dixm}.} 
 if this subgroup is reduced to the identity of $G$. 

 \bt[\cite{dixm} 16.4.8] \label{t-dix} Let $G$ be a connected, locally compact group. Then $G$ is injectable in a compact group if and only if  $G=\R^p\times K$ with $p\geq 0$ and $K$ a compact group.
\et

The second key fact that we need is due to Weil  (see \cite[p. 66]{weil}).

\bp[\cite{dixm} 13.1.8] \label{p-dix} Let $G=G_1\times G_2$ be the Cartesian product of two locally compact topological groups, and let $\Rep$ be an irreducible representation of $G$. Define the representation $\Rep'_1$ of $G_1$ as $\Rep'_1(g_1):=\Rep(g_1,e)$ and the representation $\Rep'_2$ of $G_2$ as $\Rep'_2(g_2):=\Rep(e,g_2)$. If $\Rep'_1$ and $\Rep'_2$ lie in a semisimple class of representations, then $\Rep$ is equivalent to the tensor product  $\Rep_1\otimes\Rep_2$ with $\Rep_1, \Rep_2$ irreducible representations of $G_1,G_2$, respectively.
\ep
We would need to specify what a semisimple class of representations is, see \cite[p. 65]{weil}. For our purpose, however, it is enough to recall that any class of bounded representation is semisimple (see, e.g., \cite[p. 70]{weil}).

\begin{remark} \label{l-rn}
We finally recall some elementary properties for unitary representations of $\R^p$. First recall that each irreducible unitary representation is a \emph{character}, namely, a  
representation of the type $\Chi_\xi(x):=e^{i \xi\cdot x}$ for some $\xi\in \R^p$ (see, e.g., \cite[6.1]{barut}). As a consequence we have that, for $p\geq 1$, unitary irreducible representations of $\R^p$ are not  faithful.
\end{remark}
%

\subsection{Proof of Theorem \ref{t-2}}
\label{s-proof}

 It is clear that exact controllability implies approximate controllability. The proof  that approximate controllability implies exact controllability is based on the following two results.

\bp\label{p-compact}
Let $G$ be a connected Lie subgroup of $U(n)$. If the
inclusion representation $\jmath:G\hookrightarrow U(n)$ is irreducible, then $G$ is compact.
\ep
\begin{proof}
Observe that the inclusion $\jmath:G\hookrightarrow U(n)$ is a faithful (by definition) representation of $G$ over $\C^n$, since $U(n)\subset L(\C^n)=\mbox{gl}(n,\C)$. Then, the kernel of $\jmath$ is reduced to $\{e\}$, and thus $G$ is injectable in a compact group.

Applying Theorem \ref{t-dix}, we have that $G=\R^p\times K$ with $p\geq 0$ and $K$ a compact group.
Remark that $\jmath$ is unitary, hence bounded. As already recalled, the class of bounded representations of $G$ is semisimple. Then we can apply Proposition \ref{p-dix}, that gives us two irreducible bounded representations $\Rep_1:\R^p\to L(\C^{m_1})$ and $\Rep_2:K\to L(\C^{m_2})$ such that $\jmath$ is equivalent to $\Rep_1\otimes \Rep_2$.

Since $\R^p$ is abelian and $\Rep_1$ is irreducible, then $m_{1}=1$.
%
%
Bounded irreducible (one dimensional) continuous representations of
$\R^{p}$ must be unitary. Hence $\Rep_{1}$ is a character of
$\mathbb{R}^{p}.$ 

%

%

Since $\jmath$ is faithful, then $\Rep_1$ and $\Rep_2$ are faithful too. In conclusion, $\Rep_1$ is  a faithful irreducible unitary representation of $\R^p$. Then, thanks to Remark~\ref{l-rn}, we have that $p=0$. Then $G=K$ is compact.
\end{proof}

\begin{remark}
The connectedness assumption in the statement of Proposition~\ref{p-compact} is crucial: the groups $\mbox{SE}(2,N)$ in \cite{autocitazione}  are counterexamples in the non-connected case. 
\end{remark}

\bl\label{l-irred}
Let $G$ be a subgroup of $U(n)$ such that $G z$ is dense in $S^{2n-1}$ for every $z\in S^{2n-1}$. 
Then $\jmath:G\hookrightarrow U(n)$ is an irreducible representation of $G$.
\el
\begin{proof}
Assume by contradiction that the inclusion is not irreducible, so that there exists a proper subspace $h$ of $\C^n$ which is invariant with respect to the action of $G$. Now take $z\in h\cap \Sn$ and observe that $Gz\subset h\cap \Sn$.   Thus $Gz$  is not dense, leading to a contradiction.
\end{proof}

We can now conclude the proof of Theorem~\ref{t-2}.

Let $G$ be the orbit of \eqref{gqs-lift-2}, i.e., the subgroup of $\mathscr{G}$ whose Lie algebra is generated by $\{i H(u)\mid u\in U\}$ (see Definition~\ref{d-orbit}).

Assume that  system \r{gqs-2} 
is approximately controllable.
The reachable set from a point $z\in S^{2n-1}$ for \r{gqs-2} is contained in the orbit $G z$. Hence, 
$G z$ is dense in $S^{2n-1}$ and 
Lemma~\ref{l-irred} guarantees that the inclusion 
$\jmath:G\hookrightarrow U(n)$ is an irreducible representation of $G$.
We can then apply Proposition~\ref{p-compact} and conclude that $G$ is compact.
In particular, $Gz$ is compact in $S^{2n-1}$ for every $z\in S^{2n-1}$. 
Finally, being $Gz$ dense and compact in $\Sn$, it coincides wit $\Sn$, i.e.,  system \r{gqs-2} is exactly controllable. 

Let now \r{gqs-lift-2} 
be approximately controllable. 
Hence, $G$ is dense in $\mathscr{G}$.
In particular, 
system \r{gqs-2} is also approximately controllable and, according to  the argument above, $G$ is compact. 
Hence $G=\mathscr{G}$, i.e., by Lemma~\ref{l-compact}, \r{gqs-lift-2}  is exactly controllable.
This concludes the proof of Theorem~\ref{t-2}.

\begin{remark}
If the attainable set of system \r{gqs-lift-2} is dense in any subgroup $\mathcal{G}$ of $U(n)$ which acts transitively  on $\Sn$, then
the same argument as above shows that \r{gqs-2} is exactly controllable in $\Sn$ and 
\r{gqs-lift-2} is exactly controllable in $\mathcal{G}$. 
\end{remark}

\noindent{\bf Acknowledgements:} This research has been supported by the European Research Council, ERC
StG 2009 ``GeCoMethods'', contract number 239748.

\bibliographystyle{plain}
\bibliography{biblio-spectra}

\end{document}